\documentclass[11pt]{article}
 \usepackage[letterpaper, margin=1in]{geometry}
\usepackage[numbers]{natbib} 
\usepackage{graphicx} 
 \RequirePackage[colorlinks,citecolor=blue,urlcolor=blue]{hyperref}
 \usepackage{authblk}
\usepackage{multirow,bm,xspace}

\usepackage{float,diagbox}
\usepackage{xcolor,chemarr,rotating,booktabs,fancyhdr}

\usepackage{caption}
\usepackage[font=small]{subcaption}
 
     \def\bX {\bm{X}}

\def\tbX {\tilde{\bm{X}}}
\def\bbI{\mathbf{I}}

\newcommand{\bfsym}[1]{\ensuremath{\boldsymbol{#1}}}

\def\blambda   {\bfsym{\lambda}}

\def\bxi       {\bfsym{\xi}}

\def\T {\top}

\DeclareMathOperator{\logit}{logit}

\begin{document}

\title{Tracking the Transmission Dynamics of COVID-19 with a Time-Varying Coefficient State-Space Model}

\author{{Joshua P.} {Keller}}
\author{{Tianjian Zhou}}
\author{{Andee Kaplan}}
\author{{G. Brooke Anderson}}
\author{{Wen Zhou}}
\affil{Colorado State University}

\date{}
 
  
\maketitle

\begin{abstract} 
The spread of COVID-19 has been greatly impacted by regulatory policies and behavior patterns that vary across counties, states, and countries. Population-level dynamics of COVID-19 can generally be described using a set of ordinary differential equations, but these deterministic equations are insufficient for modeling the observed case rates, which can vary due to local testing and case reporting policies and non-homogeneous behavior among individuals. To assess the impact of population mobility on the spread of COVID-19, we have developed a novel Bayesian time-varying coefficient state-space model for infectious disease transmission. The foundation of this model is a time-varying coefficient compartment model to recapitulate the dynamics among susceptible, exposed, undetected infectious, detected infectious, undetected removed, detected non-infectious, detected recovered, and detected deceased individuals. The infectiousness and detection parameters are modeled to vary by time, and the infectiousness component in the model incorporates information on multiple sources of population mobility. Along with this compartment model, a multiplicative process model is introduced to allow for deviation from the deterministic dynamics. We apply this model to observed COVID-19 cases and deaths in several US states and Colorado counties. We find that population mobility measures are highly correlated with transmission rates and can explain complicated temporal variation in infectiousness in these regions. Additionally, the inferred connections between mobility and epidemiological parameters, varying across locations, have revealed the heterogeneous effects of different policies on the dynamics of COVID-19.
\end{abstract}


\section{Introduction}
\label{sec:introduction}

Since its spread at the beginning of 2020, the COVID-19 pandemic has made clear the impact that human mobility and government policies can have on the spread of novel respiratory diseases. Throughout the world, different approaches have been taken to mitigate the spread of COVID-19, including travel bans, mask mandates, quarantine policies, and capacity restrictions. The diversity of actions taken present a unique opportunity and an urgent demand to assess the impact of these approaches quantitatively and systematically.

Statistical approaches offer opportunities to explore the relationship between
different policies and actions and disease dynamics while allowing for
variation in the outcome from both measured and unmeasured (or even
unmeasurable) factors. However, a purely marginal comparison of factors---as might be obtained using a regression-based approach---can miss important facets of the dynamics
of disease spread. This is why the foundation for most infectious disease
models is a compartmental model for the disease states, which captures these
dynamics. The traditional example of this is a compartmental ``SIR'' model that employs ordinary
differential equations to model transition from ``susceptible'' to ``infected''
to ``recovered'' states \citep{anderson1991infectious,brauercarlosfeng19,diekmann00, hethcote2000mathematics}. However, the traditional SIR model lacks the flexibility to account for the sophisticated and rapidly changing transmission dynamics of COVID-19.  
To capture the key characteristics of the COVID-19 pandemic, a number of alterations are needed for the traditional SIR model, including changes to the structure of the compartmental model, incorporation of time-varying factors that may influence the dynamics of the disease spread over the course of the pandemic, 
and introduction of stochasticity that results from limitations in model assumptions as well as noise in the observed data.

First, there are characteristics of COVID-19 spread that require an adaptation
in the structure of the compartmental model. A key characteristic of COVID-19
is that many of those infected are at their most infectious before they are
diagnosed \cite{rothe2020transmission}, with viral load typically highest at the onset of symptoms \cite{HeLauEtAl2020}.  Thus, there can be a misalignment if a compartmental model assumes people can only be infectious following diagnosis through a positive test. Further, there are many who have milder or asymptomatic cases
and are never officially diagnosed as ``infected,'' but who may still spread
the disease. The framework of a dynamic model should address these facets of
how the data (from testing) lines up with the principles of infectious spread.

Second, since the emergence of SARS CoV-2, the dynamics of COVID-19 detection
and spread have evolved, and the model should allow for an associated evolution
in some of its parameters. Public health guidance has changed throughout
2020 and 2021, with changing guidance and regulations on whether to wear masks,
travel, quarantine, gather in groups, attend in-person meetings and school,
etc., which can change the rate of transmission from those who are infected to
those who are still susceptible. Testing has improved, become more easily
accessible, and in some cases evolved to include regular testing even without
symptoms, all of which can influence the probability that someone infected with
the disease is detected.  Treatment has improved, including through the
adaptation of corticosteriods and remdesivir among patients with severe
disease and care techniques like prone positioning, use of high-flow oxygen
therapy, and intubation timing \citep{bos2020severe,boudourakis2020decreased}.
These strategies may help in reducing mortality rate of patients with severe
disease since the start of the pandemic; indeed, there is evidence of a
decrease in risk of mortality among those hospitalized for COVID-19 over 2020,
both in the US \cite{asch2020variation} and the UK \cite{dennis2020improving}.
New strains have evolved and gained prevalence during the pandemic, with
different infectivity and severity \cite{korber2020tracking, li2020impact}.
Because of these changing dynamics, parameters of the compartmental
model---including rate of detection, probability of infection given a contact
with someone infected, and rates of mortality versus recovery among the
infected---have changed over the pandemic, suggesting the need for time-varying coefficients for modeling.

Finally, there is stochasticity within the observed data compared to the
dynamics of the model. This stochasticity results both from imperfect
compliance with the assumptions of the model and from practical constraints in
collecting and reporting the data. For example, the standard compartment model assumes that a community's population is well-mixed, with an equal chance of contact among any pair of members of that population. For COVID-19, as with most infectious
diseases, this assumption is overly simplistic and requires allowance for
stochasticity, as local spikes in cases were sometimes the result of outbreaks
within an institution or organization, including significant outbreaks in
prisons, nursing homes, and meat-processing plants \cite{althouse2020superspreading, hawks2020covid,middleton2020meat}, suggesting the pandemic dynamics were in part driven by more frequent contacts within population subsets, rather than across a well-mixed community population. Further, the model assumes
homogeneity in individual transmissability, while in fact some spread is driven
by superspreaders \cite{althouse2020superspreading, adam2020clustering, lau2020characterizing, lemieux2021phylogenetic}.  Also, measurement error
is introduced through data collection. Public health officials have made an
enormous effort to collect and publish counts of cases and deaths during the
pandemic, but understandably there were occasional patterns in the data related
to data collection and reporting rather than dynamics of the virus' spread. For
example, Colorado included death counts for deaths that occurred earlier but
had not yet been reported on April 24, 2020 \cite{tabachnik2020colorado}.
Further, on weekends and holidays, reporting rates can be lower than usual,
with reporting higher following the break to incorporate the backlog
\cite{weekendcases}.

Here, we develop a time-varying  coefficient state-space model that uses a structure appropriate for COVID-19
and allows for stochasticity and measurement error in data, as well as
evolution of some model parameters over time, with the aim of investigating how
a specific factor was associated with virus spread.  This model has an
advantage over regression-based models of factors that may affect COVID case
counts over time, since under a compartmental modeling framework we are modeling the
process of disease spread, rather than correlating two time series. Further, by
incorporating elements that address time-variability in model parameters and
stochasticity inherent in the relationship between the available COVID-19 data
and the compartmental model, the model addresses limitations in a classic
SIR-style model for disease spread. 
Specifically, we introduce a multiplicative process model and a negative binomial data model to account for fluctuations in the rates of disease detection and transmission and variability in data reporting.
Focusing on retrospective estimation and inference, our model offers a framework to explore and draw inference on the effect of different human mobility behavior and related policies on the pandemic parameters in the model. It therefore meets a critical need to understand how policy choices might affect the dynamics of
spread.

As mentioned, a factor of particular interest is the influence of local mobility on COVID-19 spread
within a community. Recent studies have suggested strong ties between COVID-19
infection rates and human mobility. Initially in China, there was a strong
relationship between the number of COVID-19 cases and human mobility
\citep{kraemer2020effect}. This finding is consistent with the theory of
infectious disease spread in highly coupled metapopulations
\citep{grenfell1997meta,watts2005multiscale}. This relationship weakened after
control measures were put into place to restrict the movement of people in and
out of Wuhan province. This previous study used real-time mobility data as well as
travel history data to explore the relationship to spread of the disease. They
concluded that the drastic control measures implemented in China substantially
mitigated the spread of COVID-19. Decreased mobility was also shown to
have a protective effect against COVID-19 transmission in the USA
\citep{badr2020association}, a result that agreed with other findings that mobility inflow into a county early in the
pandemic was associated with increases in early case counts \cite{xiong2020mobile}. While
the official response in the USA to COVID-19 has been heterogeneous in terms of
lock-downs, this work showed that social distancing helps to reduce the spread
of the disease, and should remain part of personal and institutional responses
to the pandemic until a vaccine is widely adopted and should continue to be considered as a key protective public health policy in future pandemics of respiratory diseases.

While the relationship between mobility and rates of COVID-19 infection has
previously been explored \cite{kraemer2020effect,badr2020association, carteni2020mobility,glaeser2020much,iacus2020human}, analysis of this relationship has largely relied on regression-based comparisons of time series data, rather than through incorporating observed mobility within the dynamics of an epidemiological
model. Here, we develop and apply a time-varying coefficient state-space epidemiological model that
allows us to explore the relationship between mobility and COVID-19 spread in
several US states and Colorado counties. The structure of this model incorporates mobility as a factor that influences the dynamics of the
epidemic, while also accounting for variation over time in some model
parameters and stochasticity within the observed case and fatality data
compared to the model dynamics. This approach allows us to explore how mobility
influenced the dynamics of COVID-19 spread in the United States, while the
model development provides a structure that can be extended to explore how
other factors influence the dynamics of COVID-19 pandemic and spread
of other diseases in the future.

This article is organized as follows. In Section \ref{sec:math_model}, we present a deterministic compartmental model for COVID-19 with time-varying coefficients, characterized by a system of differential equations.
In Section \ref{sec:state_space_model},  by extending the compartmental model outlined in Section \ref{sec:math_model},  we develop a state-space model for COVID-19, which better accounts for stochasticity in disease spread.
We apply our proposed method to analyze county-level COVID-19 data in the U.S. state of Colorado and state-level data in the U.S. in Section \ref{sec:data_analysis}. We conclude this article with some remarks in Section \ref{sec5}. Extra results are deferred to the online supporting materials.

\section{A Time-Varying Coefficient Compartmental Model for COVID-19 Dynamics}
\label{sec:math_model}

\begin{figure}
\centering
\includegraphics[scale=0.8]{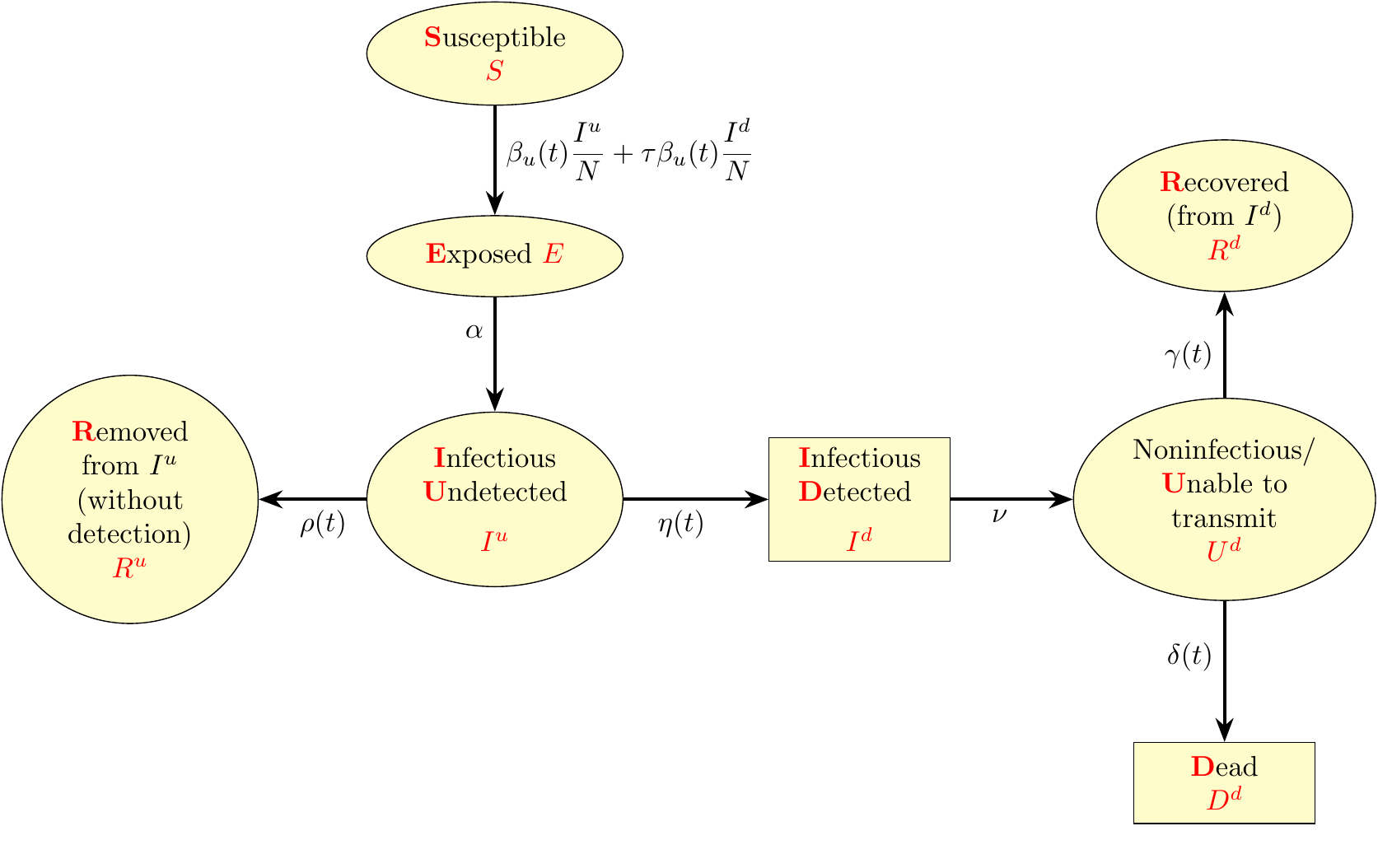}
\caption{The conceptual dynamics for our eight-compartment time-varying coefficient dynamical model. Square nodes represent observed states and elliptical nodes represent unobserved states. Compartment transitions are labeled with their rates.}
\label{fig0}
\end{figure}

Given its flexibility in structure and natural connection with dynamical systems, the compartmental modeling approach has been extensively employed in epidemiology \citep{brauercarlosfeng19,diekmann00,watts2005multiscale,Cauchemez2004,Phenyo06} and pharmacokinetics \citep{edmunds1999,machera2006}.
This approach has been widely adopted to understand the dynamics of COVID-19  \citep{HeLauEtAl2020,hao2020reconstruction,LiPeiEtAl2020}. 
We first outline a basic compartmental model for recapitulating the transmission dynamics of COVID-19, which is characterized by a system of differential equations. It will serve as the cornerstone for our statistical model in Section \ref{sec:state_space_model}.
We partition the population in a region (which can be a county, state, or country) into the following eight compartments, each representing a specific stage of COVID-19:
\begin{enumerate}
\item Susceptible individuals ($S$): those who have not been infected by the disease and are at risk of becoming infected;
\item Exposed individuals ($E$): those who have been exposed to the disease but are not yet capable of infecting others;
\item Undetected infectious individuals ($I^u$): those who have the disease, are able to infect others, but have not yet been detected as having the disease;
\item Detected infectious individuals ($I^d$): those who have the disease, are able to infect others, and have been detected as having the disease;
\item Individuals removed from $I^u$ without being detected ($R^u$): those who had the disease but
are then removed from the possibility of being infected again or spreading the disease, either through recovery or death, without ever being diagnosed with the disease;
\item Individuals removed from $I^d$ ($U^d$): those who had the disease, were diagnosed, but are then removed from the possibility of being infected again or spreading the disease;
\item Individuals recovered from $U^d$ ($R^d$): those who have been through the $U^d$ state and are eventually recovered from the disease (in terms of no more symptoms);
\item Individuals died from $U^d$ ($D^d$): those who have been through the $U^d$ state and are eventually deceased.
\end{enumerate}
The eight compartments, together with the possible transitions among compartments, are illustrated in Figure \ref{fig0} and are designed to capture important features of the dynamics of COVID-19.
First, it has been well recognized that asymptomatic and pre-symptomatic individuals, or those who are symptomatic but have not yet been diagnosed with the disease, contribute significantly to the spread of COVID-19 \citep{rothe2020transmission,HeLauEtAl2020}. These individuals are modeled through the $I^u$ compartment in the proposed partition. In addition, recent research \citep{HeLauEtAl2020, wolfel2020virological} indicates that the infectiousness of an infected individual declines quickly within a week of symptom onset. 
This suggests that an infected individual may no longer be capable of infecting others before the complete disappearance of symptoms or death.
To reflect this, the proposed model specifically allows the individuals in $I^d$ to first go through the $U^d$ state before recovery or death.
Since no data on the recovery or death of the $I^u$ individuals are available, it is unnecessary to include additional compartments for individuals who recovered or died from the $R^u$ state, and neither will this affect the essence of transmission dynamics.

We use the notation of a compartment followed by a time index $t$ to denote the size of the compartment at a specific time point. For example, $S(t)$ denotes the number of susceptible individuals at time $t$. The transmission of COVID-19 can be characterized by the flow of individuals through these compartments  over time, given by the following system of differential equations with time-varying coefficients:
 \allowdisplaybreaks
 \begin{align}
\displaystyle \frac{dS(t)}{dt}& = -\beta_u(t)N^{-1}(t)\{I^u(t) + \tau I^d(t)\}S(t), \label{e21} \\
\displaystyle \frac{dE(t)}{dt}& =  \beta_u(t)N^{-1}(t)\{I^u(t) + \tau I^d(t)\}S(t)  - \alpha E(t), \label{e22} \\
\displaystyle \frac{dI^u(t)}{dt}& = \alpha E(t) - \left\{\eta(t) + \rho(t) \right\}I^u(t),  \label{e23}\\
\displaystyle \frac{dR^u(t)}{dt}& = {\rho(t)}I^u(t),  \label{e26}\\
\displaystyle \frac{dI^d(t)}{dt}& =\eta(t)I^u(t) - \nu I^d(t), \label{e24} \\
\displaystyle \frac{dU^d(t)}{dt}& = \nu I^d(t) -\left\{\gamma(t) + \delta(t)\right\}U^d(t), \label{e25} \\
\displaystyle \frac{dR^d(t)}{dt}& = \gamma(t) U^d(t), \label{e76} \\
\displaystyle \frac{dD^d(t)}{dt}& = \delta(t) U^d(t)\label{e28}.
\end{align}
Here, $N(t)=S(t)+E(t)+I^u(t)+I^d(t)+U^d(t)$ denotes the population of active individuals.
A simplification in model \eqref{e21}-\eqref{e28} is that we ignore births and deaths, an approximation that is appropriate for a fast-spreading pandemic like COVID-19.
For simplicity, we also do not consider immigration and emigration.
Given the initial condition (i.e., the initial sizes of the compartments) and parameter values, the trajectory of the epidemic process can be deterministically obtained by solving the system of differential equations.
In Sections \ref{sec:state_space} and \ref{sec:data}, we will discuss how to allow for additional flexibility and stochasticity in the time-varying coefficient compartmental model \eqref{e21}-\eqref{e28}, and how to link the observed data on COVID-19 cases and deaths to the epidemic process.
The initial conditions of the model are not trivial and will be discussed in Section \ref{sec:initial_condition}.

The time-varying coefficients in \eqref{e21}-\eqref{e28} bring extra flexibility to the compartmental model to capture the complex and rapidly-varying dynamics of COVID-19. Specifically, $\beta_u(t)$ models the {\it disease transmission rate} between the undetected infectious and susceptible individuals at time $t$.
The rate $\beta_u(t)$ can be understood as the number of effective contacts (i.e., contacts that are sufficient for disease transmission) made by an average member of the undetected infectious individuals per unit time. 
The probability of each of these contacts being with a susceptible individual is $S(t)/N(t)$.
Therefore, $I^u(t)$ undetected infectious individuals lead to a rate of new infections $\beta_u(t) N(t)^{-1} I^u(t) S(t) $. 
For the detected infectious individuals, the rate of effective contacts is reduced by a factor of $\tau \in [0,1]$, since those diagnosed with COVID-19 are likely to have reduced contact with others due to potential (self-)quarantine or hospitalization, leading to a rate of new infections $\tau \beta_u(t) N(t)^{-1} I^d(t) S(t)$. To incorporate additional covariate information such as mobility and changes in policy, a detailed model of $\beta_u(t)$ will be provided 
in Section \ref{sec:transmission_rate}.

After being infected, a susceptible individual first goes through an ``exposed'' state, meaning that the individual has been exposed to the disease but is not immediately able to infect others. The exposed individuals enter the undetected infectious state at a rate of $\alpha$, and $\alpha^{-1}$ naturally represents the {\it latent period}, which is the time interval between when an individual is exposed to the disease and when the individual becomes capable of infecting other susceptible individuals. 
Then, the undetected infectious individuals are either diagnosed with the disease at a detection rate of $\eta(t)$ or removed from the infectious state at a rate of $\rho(t)$ without ever being diagnosed.
Similar to $\beta_u(t)$, the detection rate $\eta(t)$ is expected to vary along with the development of COVID-19 pandemic.  Estimation of $\eta(t)$ plays an indispensable role in understanding the epidemiological mechanics of COVID-19. 
In our model, we further assume that an infectious individual always needs to go through the ``undetected'' state before the individual is detected as having the disease. 
It is possible that an exposed individual can be directly diagnosed with the disease through contact tracing, but we treat this as a special case that the individual spends zero time at the undetected infectious state. 
Infectious individuals that have been detected become non-infectious at rate $\nu$.  From this state, individuals are finally removed either through recovery at a  rate of $\gamma(t)$ or due to decease at a rate of  $\delta(t)$.
Naturally, deceased individuals do not contribute to the dynamics of disease transmission. 
We also assume that recovery from COVID-19 confers immunity to reinfection, and thus the recovered individuals can neither spread the disease nor be infected again. This is a simplifying assumption motivated by the rate of reinfection being so low that it does not meaningfully alter the transmission dynamics.
In Section \ref{sec:param_explain} we discuss the modeling and interpretation of each of these parameters in greater detail.
 
For epidemiological models, the reproductive ratio is a fundamental quantity to track the pandemic and it serves as a threshold that predicts the spread of an infection \cite{brauercarlosfeng19,diekmann00}. For models that are more sophisticated than the simple SIR model, the reproductive ratio is usually computed based on the {\it equilibrium reproduction number}, which characterizes the persistence of the pandemic. For the proposed model \eqref{e21}-\eqref{e28}, the equilibrium reproduction number is
\begin{equation}
\mathcal{R}_{0,e}(t)=\frac{\beta_u(t)}{\eta(t)+\rho(t)}\left\{1+\eta(t)\frac{\tau}{\nu}\right\},
\label{r0e1b}
\end{equation} 
 which is derived using the endemic equilibrium argument \citep{HSW05} in
   Section A in the supplementary file.
 If $\mathcal{R}_{0,e}(t) < 1$ after time $t^*$,  the number of infectious individuals will decrease after time $t^*$ and lead to the disease-free equilibrium.  Therefore, an $\mathcal{R}_{0,e}(t) < 1$ indicates containment of the disease at time $t$. On the contrary, if $\mathcal{R}_{0,e}(t) >1$ after time $t^*$, the pandemic will persist with a nontrivial equilibrium \citep{HSW05} and the larger the equilibrium reproduction number, the larger the population of infectious individuals at the equilibrium will be.  Due to its important role in characterizing disease spread, inference on $\mathcal{R}_{0,e}(t)$ is one of our major interests.  In addition, differentiating $\mathcal{R}_{0,e}(t)$ against $\eta(t)$, we observe that $\mathcal{R}_{0,e}(t)$ is monotonically decreasing in $\eta(t)$ if and only if $\rho(t) <\nu/\tau$.  Assume that the removal rates of the undetected and detected infectious individuals are similar. If $\tau > 1$, the condition $\rho(t) <\nu/\tau$ might not hold, meaning that more efficient detection would lead to more disease transmission.  This is inconsistent with the known facts about the COVID-19 pandemic. By restricting $\tau \leq 1$ and modeling the relationship between $\rho(t)$ and $\nu$ (accomplished via prior specification, see Section \ref{sec:param_explain}), it is guaranteed that $\rho(t) <\nu/\tau$.  As $\eta(t)\to 0$, the difficulty of detection increases  and  $\mathcal{R}_{0,e}(t)$ monotonically converges to $\beta_u(t)/\rho(t)$. That is, the pandemic dynamics are  dominated by the  undetected infectious individuals.  On the other hand, as $\eta(t)$  increases,  the deviation between $\mathcal{R}_{0,e}(t)$  and $\beta_u(t)\tau/\nu$   shrinks. That is, the pandemic dynamics will be governed more by the detected infectious individuals.

\section{State-Space Model for COVID-19 with Time-Varying  Coefficients}
\label{sec:state_space_model}

While the compartmental model in Section \ref{sec:math_model}  describes the underlying deterministic and smooth disease dynamics, there is stochasticity in the actual disease spread. This is due to factors such as non-uniform mixing of the population and heterogeneity in day-to-day activity patterns.
Additionally, data on COVID-19 cases and deaths are always observed with substantial statistical noise, such as reporting error and reporting delay, which cannot  be easily captured by the differential equations.
To  mitigate these issues, we cast the compartmental model in Section \ref{sec:math_model} in a state-space modeling framework.
Following an approach similar to Dukic \textit{et al.}\cite{DukicLopesEtAl2012} and Osthus \textit{et al.}\cite{OsthusHickmannEtAl2017}, 
we construct a {\it process model} which allows the epidemic process to stochastically deviate from the solution given by the compartmental model \eqref{e21}-\eqref{e28}.
Moreover,  we build a {\it data model} for the observed data, which further takes into account the measurement error of COVID-19 cases and death counts.

\subsection{Process Model}
\label{sec:state_space}

Since COVID-19 data are reported daily, it is natural to consider a discretized version of the compartmental model in Section \ref{sec:math_model} with a time step of 1 day.
Denote $\bX(t) = [S(t), E(t), I^u(t), R^u(t),$ $I^d(t), U^d(t), R^d(t), D^d(t)]^{\T}$ the true but unobservable populations of the eight compartments on day $t = 0, 1, 2, \ldots$.
Define $\tbX(t) = [\tilde{S}(t), \ldots, \tilde{D}^d(t)]^{\T}$ as the solution to differential equations \eqref{e21}-\eqref{e28}, 
starting from the state of the prior day, $\bX(t-1)$.

Denote by $\mu^C(t)$ the unobservable number of individuals diagnosed with COVID-19 on day $t$. Similarly, let $\mu^D(t)$ denote the unobservable number of individuals who died from COVID-19 on day $t$.
In terms of  states $\bX(t)$ and $\bX(t-1)$, 
 $\mu^C(t)$ represents the number of individuals who moved from $I^u(t - 1)$ to $I^d(t)$, and $\mu^D(t)$ represents the number of individuals who moved from $U^d(t - 1)$ to $D^d(t)$. As discussed before, it is unrealistic to assume that the true yet unobservable numbers of the diagnosed and reported deceased individuals  perfectly agree with the solution to the compartmental model, especially for the complicated pandemics such as COVID-19. To introduce additional flexibility, motivated by Davis and Wu\cite{DavisWu2009a}, we model $\mu^C(t)$ and $\mu^D(t)$ via  multiplicative processes
\begin{equation}
\mu^C(t) = \tilde{\mu}^C(t)\epsilon^C(t) ~ \text{and}~
\mu^D(t) = \tilde{\mu}^D(t)\epsilon^D(t),
\label{state_m}
\end{equation}
where $\tilde{\mu}^C(t)$ and $\tilde{\mu}^D(t)$ refer to the numbers of new detections and deaths on day $t$, according to the solution $\tbX(t)$ to our compartmental model.
For simplicity, we model the multiplicative processes $\epsilon^C(t)$ and $\epsilon^D(t)$ as  random variables independent in $t$ with
\begin{align*}
\epsilon^C(t)\sim {\rm Gamma}(1/(v^C)^2, 1/(v^C)^2)  ~ \text{and}~
\epsilon^D(t) \sim {\rm Gamma}(1/(v^D)^2, 1/(v^D)^2),
\end{align*}
 so that ${\rm E}\{\epsilon^C(t)\} =1$, ${\rm Var}\{\epsilon^C(t)\} = (v^C)^2$,   ${\rm E}\{\epsilon^D(t)\} =1$, and ${\rm Var}\{\epsilon^D(t)\} = (v^D)^2$.
The unit mean assumption, which mimics the zero mean assumption for the additive models, leads to ${\rm E}\{\mu^C(t) \mid \tilde{\mu}^C(t)\} = \tilde{\mu}^C(t)$ and   ${\rm E}\{\mu^D(t) \mid \tilde{\mu}^D(t)\} = \tilde{\mu}^D(t)$. Compared to the traditional additive models, the multiplicative process model \eqref{state_m} is more natural to introduce deviations between the nonnegative solution to differential equation models and unobservable process with nonnegative observable space. In summary, the underlying epidemic process is centered at the solution to  differential equations  \eqref{e21}-\eqref{e28} but is also allowed to be stochastically different from the  deterministic process specified by the compartmental model.

The solution $\tbX(t)$ to the differential equations \eqref{e21}-\eqref{e28} is not available in closed form. 
A numerical approximation method, such as the Runge-Kutta solver, must be employed.
Here we use a simple and computationally efficient first-order difference calculation, which does not compromise much the model performance. From \eqref{e24}, a first-order difference update to the states means that $\tilde{I}^d(t) = (1 - \nu)I^d(t-1) + \eta(t)I^u(t-1) $. 
Thus, $\tilde{\mu}^C(t) = \eta(t)I^u(t-1)$, and $\mu^C(t) = \eta(t) \epsilon^C(t) I^u(t-1)$.
We assume that $\epsilon^C(t)$ in \eqref{state_m} only affects the transition from state $I^u$ to state $I^d$. Thus, we have
\begin{align*}
I^u(t) - I^u(t-1) &=  \alpha E(t-1) - \{\eta(t-1) \epsilon^C(t-1) + \rho(t) \}I^u(t-1)\\
I^d(t) - I^d(t-1) &=  \eta(t-1) \epsilon^C(t-1) I^u(t-1) - \nu I^d(t-1).
\end{align*}
In essence, the error process $\epsilon^C(t)$ therefore offers additional flexibility  in the detection rate $\eta(t)$ from the compartmental model.
Accordingly, the empirical calculation of $\mathcal{R}_{0,e}(t)$ will be adjusted by replacing $\eta(t)$ in \eqref{r0e1b} with $\eta(t) \epsilon^C(t)$. The error process $\epsilon^D(t)$ plays a similar role as $\epsilon^C(t)$ in the disease transmission, and we assume that they only affect the transition from $U^d$ to $D^d$. Specifically,
\begin{align*}
U^d(t) - U^d(t-1) &=  \nu I^d(t-1) -\{\gamma(t-1) + \delta(t-1) \epsilon^D(t-1) \}U^d(t-1), \\
D^d(t) - D^d(t-1) &=  \delta(t-1) \epsilon^D(t-1) U^d(t-1).
\end{align*}
As both $U^d$ and $D^d$ individuals are unable to spread the disease,  $\epsilon^D(t)$ does not influence the equilibrium reproduction number.

Lastly, based on our assumptions, the error terms do not have an impact on the states $S(t)$, $E(t)$, $R^u(t)$, and $R^d(t)$, and these states are consistent with the corresponding  solutions to the compartmental model $\tilde{S}(t)$, $\tilde{E}(t)$, $\tilde{R}^u(t)$, and $\tilde{R}^d(t)$, calculated using the first-order approximation.

\subsection{Data Model}
\label{sec:data}
Let $Y^C(t)$ be the number of newly reported positive cases on day $t$, and denote $Y^D(t)$ the  number of newly reported deaths on day $t$, where $t \in \{0, 1, 2, \ldots \}$.
In the event that only cumulative numbers of cases and deaths are available, the values of $Y^C(t)$ and $Y^D(t)$ can be computed using a first order difference. 
We assume negative binomial models for the daily reported case and death counts,
\begin{align*}
Y^C(t) \sim \text{NegBinom}( \mu^C(t), \phi_C ) ~\text{and}~
Y^D(t) \sim \text{NegBinom}(\mu^D(t), \phi_D ),
\end{align*}
where $\mathrm{E}\{Y^C(t)\} = \mu^C(t)$ and $\mathrm{Var}\{Y^C(t)\} = \mu^C(t)\{1  + \mu^C(t)\phi_C^{-1}\}$, respectively. Parameter $\phi_C$ allows for overdispersion, relative to a Poisson distribution, and grants more flexibility to the model for fitting the COVID-19 data \cite{LoroMingione21}.
We place a Gamma prior on $\phi_C$, $\phi_C \sim \text{Ga}(50, 100)$, while $\phi_D$ is treated similarly.

\subsection{Model for the Initial Condition} 
\label{sec:initial_condition}

The initial condition of the infection dynamics refers to the initial population sizes of the compartments. 
The calendar time corresponding to $t=0$ can be chosen for each modeling context. We assume that at $t = 0$, the total number of removed individuals is 0, i.e., $R^u(0) = R^d(0) = U^d(0) = D^d(0) = 0$. The number of detected infectious individuals at time $0$, $I^d(0)$, is observed and assumed to be non-zero. In the analyses of Section~\ref{sec:data_analysis}, we choose $t=0$ to be a time early in the pandemic when there are multiple detected cases.

 The numbers of susceptible, exposed, and undetected infectious individuals at time $0$ are not observed and need to be estimated. We assume the initial population size of the exposed and undetected infectious individuals  is $\kappa$ times the initial size of the detected infectious individuals,
\begin{align*}
E(0) + I^u(0) = \kappa  I^d(0).
\end{align*}
We place a gamma distribution prior on $\kappa$, $\kappa\sim \text{Ga}(25, 5)$, with a prior mean of 5. This choice is based on the findings by Li \textit{et al.}\cite{LiPeiEtAl2020} that 86\% of all infections were undocumented at the beginning of the epidemic in China.

Next, we place a uniform distribution prior on the proportion of exposed individuals among $E(0) + I^u(0)$. That is, we assume $E(0) / [E(0) + I^u(0)] \sim \text{Unif}(0, 1).$
Lastly, we have $S(0) = N(0) - E(0) - I^u(0) - I^d(0)$, where $N(0)$ is the (known) population size.

\subsection{Model for the Time-Varying Disease Transmission Rate with Covariates}
\label{sec:transmission_rate}

We now turn to the modeling of the epidemiological parameters.
We start with the model construction for the time-varying disease transmission rate between the undetected infectious and susceptible individuals, $\beta_u(t)$,
which is an important parameter that characterizes the speed of disease transmission.
We model the transmission rate as time-varying to account for changes in COVID infection rates due to human behavior \citep{badr2020association,carteni2020mobility}, enactment of government policies \citep{kraemer2020effect,TianTanEtAl2021}, evolution of new viral strains, and other time-sensitive factors. Since $\beta_u(t) > 0$, it is easier to work with the log transformation of $\beta_u(t)$. A simple and flexible way of modeling $\log\{\beta_u(t)\}$ is through temporal splines,
\begin{equation}
\log\{\beta_u(t)\} = \zeta_{\beta} + \blambda_{\beta}(t)^{\top} \bxi_{\beta},
\label{betau}
\end{equation} 
where $\zeta_{\beta}$ is the regression intercept term, $\blambda_{\beta}(t)$ is a vector of basis functions evaluated at time $t$ (excluding the intercept), and $\bxi_{\beta}$ is the vector of regression coefficients. We primarily consider a piecewise-linear spline for $\blambda_{\beta}(t)$ with knots selected by quantiles, but other forms (piecewise constant, natural cubic splines, etc.) are possible in this framework. The intercept term and regression coefficients are further modeled through normal priors, $\zeta_{\beta} \sim N(0, 1)$ and  $\bxi_{\beta} \sim N(\bm 0, \sigma_{\beta}^2 \bbI)$
with $\sigma_{\beta} \sim N_+(0.5, 0.1^2)$.
Here, $N_+(0.5, 0.1^2)$ denotes a normal distribution with mean 0.5 and standard deviation 0.1 restricted to $(0, \infty)$.

The framework in \eqref{betau} makes it straightforward to incorporate additional covariates for modeling $\beta_u(t)$.
Let $\theta_1(t), \ldots, \theta_L(t)$ denote $L$ covariates.
Motivated by the hybrid of single index models
and additive models\cite{ma2017siam}, 
we further generalize \eqref{betau} as
\begin{equation}
\log\{\beta_u(t)\} = \zeta_{\beta} + \blambda_{\beta}(t)^{\top} \bxi_{\beta} + \sum_{\ell=1}^L \theta_{\ell}(t)  \xi_{\beta \theta, \ell}.
\label{betau_s}
\end{equation} 
In Section~\ref{sec:data_analysis}, we will include the information on human mobility via the terms $\theta_{\ell}(t)$.

The transmission rate between the detected infectious and susceptible individuals is assumed to be reduced by a factor of $\tau \in [0,1]$ due to potential quarantine and hospitalization.
We place a uniform distribution prior on $\tau$, $\tau \sim \text{Unif}(0, 1)$.

\subsection{Model for the Other Epidemiological Parameters}
\label{sec:param_explain}

In this section, we will detail the model and prior specification for the remaining epidemiological parameters. 
Most parameters in model  \eqref{e21}-\eqref{e28} have practical implications corresponding to clinical characteristics of COVID-19.
Therefore, we elicit informative priors for these parameters by summarizing the findings in the literature \cite{HeLauEtAl2020,hao2020reconstruction,LiPeiEtAl2020,deng2020estimation, ferretti2020timing, ferretti2020quantifying, Jiang2021,  LiGuanEtAl2020,  VerityEtAl2020}.
We acknowledge that there is not yet a consensus on all clinical characteristics of COVID-19. For example, Li et al \citep{LiGuanEtAl2020} estimated a mean incubation period of $5.2$ days, while the same quantity was estimated to be $7.75$ and $9.1$ days in Jiang et al \citep{Jiang2021} and Deng et al \citep{deng2020estimation}, respectively.
As a result, our priors are chosen in consistent with the majority of the literature. 
Our general methodology works for any choices of priors.
We also note that the timing of COVID-19 events is highly variable at the individual level, but the parameters in \eqref{e21}-\eqref{e28} are defined at the population level, which are less variable.
The informative priors improve the interpretability of the results by assigning larger prior mass to the parameters around clinically meaningful values.
In addition, since our model is an eight-compartment model with only two compartments observed, some model parameters are nonidentifiable.  Such nonidentifiablility is mitigated through the use of informative priors \citep{neath1997efficacy, wechsler2013bayesian}.

In the following paragraphs, we detail the interpretation and modeling choices for the remaining parameters, which are summarized in Tables~\ref{tbl:prior} and \ref{tbl:prior_clinical}.
\begin{table}
\caption{Model parameters, interpretations, and prior distributions.} 
\small
\label{tbl:prior}
\begin{center}
\begin{tabular}{lll}
\toprule
Parameter & Interpretation & Prior   \\
\midrule
$E(0)$ &  Initial number of exposed individuals & $\dfrac{E(0)}{E(0) + I^u(0)}\sim \text{Unif}(0, 1)$ \\
$\tau$ & Reduction in transmission rate for detected infectious individuals & $\text{Unif}(0, 1)$ \\
\midrule
$\phi_C$ & Overdispersion parameter for daily cases & $\text{Ga}(50, 100)$ \\
$\phi_D$ & Overdispersion parameter for daily deaths & $\text{Ga}(50, 100)$ \\
\midrule
$\zeta_{\beta}$ & Intercept for disease transmission rate (log scale) & $N(0, 1)$ \\
$\bxi_{\beta}$, $\bxi_{\beta \theta}$ & Regression coefficients for disease transmission rate (log scale) & $N(\bm 0, \sigma_{\beta}^2 \bbI)$ \\
$\sigma_{\beta}$ & Standard deviation of regression coefficients for disease transmission rate & $N_{+}(0.5, 0.1^2)$ \\
$a_{\psi}$ &  Stabilized detection fraction & $\text{Be}(55.5, 18.5)$ \\
$b_{\psi}$ & $a_{\psi} - b_{\psi}$ is the initial detection fraction & $b_{\psi} / a_{\psi} \sim \text{Unif}(0, 1)$\\
$c_{\psi}$ & Speed of detection fraction increase & $\text{Ga}(5, 100)$ \\
$\zeta_{\omega}$ & Intercept for death fraction (logit scale) & $N(-1, 1)$ \\
$\bxi_{\omega}$ & Regression coefficients for death fraction (logit scale) & $N(\bm 0, \sigma_{\omega}^2 \bbI)$ \\
$\sigma_{\omega}$ & Standard deviation of regression coefficients for death fraction & $N_{+}(0.5, 0.1^2)$ \\
\bottomrule
\end{tabular}
\end{center}
\end{table}

\begin{table}
\caption{Model parameters with informative priors elicited based on clinical characteristics of COVID-19.} 
\small
\label{tbl:prior_clinical}
\begin{center}
\begin{tabular}{lll}
\toprule
Param. & Interpretation & Prior \\
\midrule
$\kappa$ & Multiplicative factor of initial exposed and undetected individuals relative to detected & $\text{Ga}(25, 5)$  \\
$\alpha$ & Inverse of latent period & $\text{Be}(31.5, 58.5)$ \\
$\eta_0$ & Inverse of infectious period before detection (for detected individuals) & $\text{Be}(32.4, 58.4)$ \\
$\nu$ &  Inverse of infectious period after detection (for detected individuals) & $\text{Be}(6.9, 41.1)$ \\
$\rho_0$ &  Inverse of whole infectious period (for undetected individuals)  & $\left( \eta_0^{-1} + \nu^{-1} \right)^{-1}$ \\
$\gamma_0$ & Inverse of time to recovery after the end of infectious period (for detected individuals) & $\text{Be}(21.5, 431.0)$  \\ 
$\delta_0$ & Inverse of time to death after the end of infectious period (for detected individuals) & $\text{Be}(47.3, 615.0)$  \\ 
\bottomrule
\end{tabular}
\end{center}
\end{table}

\begin{figure}
\begin{center}
\includegraphics[width = .95\textwidth]{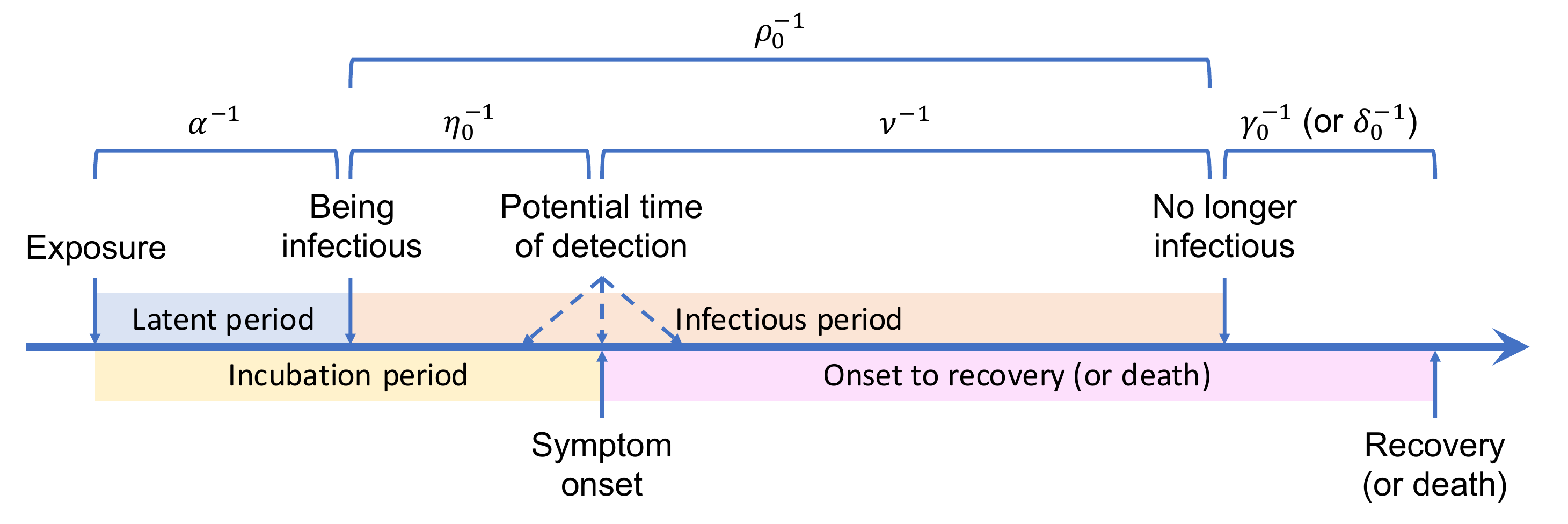}
\end{center}
\caption{Natural history of COVID-19, definitions of terms, and corresponding parameters. }
\label{fig:covid_natural_history}
\end{figure}

\paragraph{Latent period.} The parameter $\alpha$ denotes the rate of exposed individuals becoming infectious, and $\alpha^{-1}$ represents the latent/pre-infectious period, i.e., the time period between exposure to the disease and being able to infect others.
Note that the latent period in our paper is different from the incubation period, where the latter refers to the time period between exposure and symptom onset (see Figure \ref{fig:covid_natural_history}).
For COVID-19, it is well recognized that patients usually become infectious before the onset of symptoms \citep{HeLauEtAl2020}.
We place a beta distribution prior on $\alpha$, $\alpha \sim \text{Be}(31.5, 58.5)$, with a prior mean of  $1/2.9$ and standard deviation of $0.05$.

\paragraph{Detection rate.} 
The parameter $\eta(t)$ represents the rate of undetected infectious individuals being detected as having the disease at time $t$.
We model $\eta(t)$ by $\eta(t) = \psi(t)  \eta_0,$
where $\psi(t)$ represents the proportion of undetected infectious individuals on day $t$ that are eventually diagnosed with the disease, and $\eta_0$ is a parameter such that $\eta_0^{-1}$ represents the time interval between when an individual is capable of infecting others and when the individual is detected as having the disease.
In practice,  detection may occur after the onset of symptoms (e.g., a person feels sick a day or two and then goes to get a test) or earlier than the start of symptoms (e.g., a group like a college athletic organization getting mandatory regular tests that identify pre-symptomatic or asymptomatic cases).
Without additional information, we assume that the average time to detection for undetected infectious individuals is similar to the time to symptoms onset. In other words, on average, undetected infectious individuals (who are eventually diagnosed) are diagnosed with the disease when their symptoms start to appear.
Therefore, $\alpha^{-1} + \eta_0^{-1}$ is roughly equal to the incubation period.
We place a beta distribution prior on $\eta_0$, $\eta_0 \sim \text{Beta}(32.4, 58.4)$, with a prior mean of  $1/2.8$ and standard deviation of $0.05$.
Next, we model $\psi(t)$ with a three-parameter curve,
\begin{align}
\psi(t) = a_{\psi} - b_{\psi}  \exp(-c_{\psi} t),
\label{psit}
\end{align}
where $0 < b_{\psi} < a_{\psi} < 1$, and $c_{\psi} > 0$. As a result, $\psi(t)$ starts from $a_{\psi} - b_{\psi}$ at time $t = 0$,  monotonically increases with increasing $t$, and converges to $a_{\psi}$ as $t \rightarrow \infty$.
The underlying assumption is that at the beginning of the pandemic, due to the limited testing capacity, the proportion of detection starts from a low level $a_{\psi} - b_{\psi}$.
As time progresses, testing capacity increases monotonically, and eventually, anyone who wants a test can get it, making the proportion of detection stabilize at $a_{\psi}$.
The parameter $c_{\psi}$ characterizes the speed of testing capacity increase.
We assume $a_{\psi} \sim \text{Be}(55.5, 18.5)$ with a prior mean of  $0.75$ and standard deviation of $0.05$.
Further, we model the ratio between $b_{\psi}$ and $a_{\psi}$ and place a uniform distribution prior on it, $b_{\psi} / a_{\psi} \sim \text{Unif}(0, 1)$. Lastly, we assume $c_{\psi} \sim \text{Ga}(5, 100)$.

\paragraph{Infectious period.} 
The parameter $\nu$ denotes the rate of detected infectious individuals becoming noninfectious, and $\nu^{-1}$ represents the time interval between disease detection and when an individual is no longer able to infect others. 
We place a beta distribution prior on $\nu$, $\nu \sim \text{Beta}(6.9, 41.1)$,
with a prior mean of  $1/7$ and standard deviation of $0.05$. 

The parameter $\rho(t)$ denotes the rate of undetected infectious individuals becoming noninfectious without ever being detected as having the disease. We have $\rho(t) = \{1 - \psi(t)\} \rho_0.$
Here, $1 - \psi(t)$ is the proportion of undetected infectious individuals on day $t$ that are never diagnosed with the disease, and $\gamma_0^{-1}$  is the time interval during which a never-detected infectious individual is capable of infecting others, i.e., the whole infectious period. By the definitions of $\eta_0$ and $\nu$, we have $\rho_0^{-1} = \eta_0^{-1} + \nu^{-1}.$
In other words, the whole infectious period equals to the time interval between when an individual is capable of infecting others and when the individual is diagnosed with the disease, plus the time interval between diagnosis and when the individual is no longer able to infect others. Based on such a relationship, it is guaranteed that $\rho(t) < \nu$, ensuring that $\mathcal{R}_{0,e}(t)$ is a monotonically decreasing function of $\eta(t)$ (see Equation \eqref{r0e1b}).

\paragraph{Recovery and death rates.} 

The parameters $\gamma(t)$ and $\delta(t)$ are the rates of recovery and death for detected but no-longer-infectious individuals at time $t$, respectively.
We have 
\begin{equation}
\gamma(t) = \{1 - \omega(t)\}\gamma_0 ~\text{and}~ 
\delta(t) = \omega(t)\delta_0,
\label{omega}
\end{equation}
where $\omega(t)$ represents the fraction of deaths among detected but no-longer-infectious individuals at time $t$, and accordingly, $1 - \omega(t)$ represents the corresponding fraction of recoveries. By constructing $\gamma(t)$ and $\delta(t)$ as a function of the time-invariant $\delta_0$ and $\gamma_0$ and the time-varying $\omega(t)$ in \eqref{omega}, we can easily estimate a time-varying recovery rate and time-varying death rate without introducing additional identifiability concerns.

We place beta distribution priors on $\gamma_0$ and $\delta_0$. Specifically, $\gamma_0 \sim \text{Be}(21.5, 431.0)$ with a prior mean of  $1/21$ and standard deviation of $0.01$, and $\delta_0  \sim \text{Be}(47.3, 615.0)$ with a prior mean of  $1/14$ and standard deviation of $0.01$.
The fraction of deaths $\omega(t) \in (0, 1)$, and therefore, we consider the logit transformation of $\omega(t)$, $\logit [\omega(t)] = \log \left\{ \omega(t) / [1 - \omega(t)] \right \}$. We model the transformed $\omega(t)$ with a B-spline,
\begin{equation}
\logit \{\omega(t)\} =  \zeta_{\omega} + \blambda_{\omega}(t)^{\top} \bxi_{\omega}.
\label{omega}
\end{equation} 
Here, $\zeta_{\omega}$ is the intercept term, $\blambda_{\omega}(t)$ is a vector of basis functions evaluated at time $t$ (excluding the intercept), and $\bxi_{\omega}$ is the vector of regression coefficients. 
The intercept term and regression coefficients are given normal distribution priors, $\zeta_{\omega} \sim N(-1, 1)$ and  $\bxi_{\omega} \sim N(\bm 0, \sigma_{\omega}^2 \bbI)$
with $\sigma_{\omega} \sim N_+(0.5, 0.1^2)$.

\subsection{Implementation}

We use the Stan modeling framework to sample from the posterior distribution of all parameters \cite{StanDevelopmentTeam2017}. Stan uses a Hamiltonian Monte Carlo procedure to efficiently sample from the posterior distribution. A major advantage to using the Stan framework is that it can easily accommodate non-conjugate and truncated prior distributions, which allows us for more flexibility in choosing a prior distribution that is scientifically relevant for each parameter.

\section{Analysis of the COVID-19 Data}
\label{sec:data_analysis}

\subsection{Data Collection}
\begin{figure}
\begin{center}
\includegraphics[width = 0.8\textwidth]{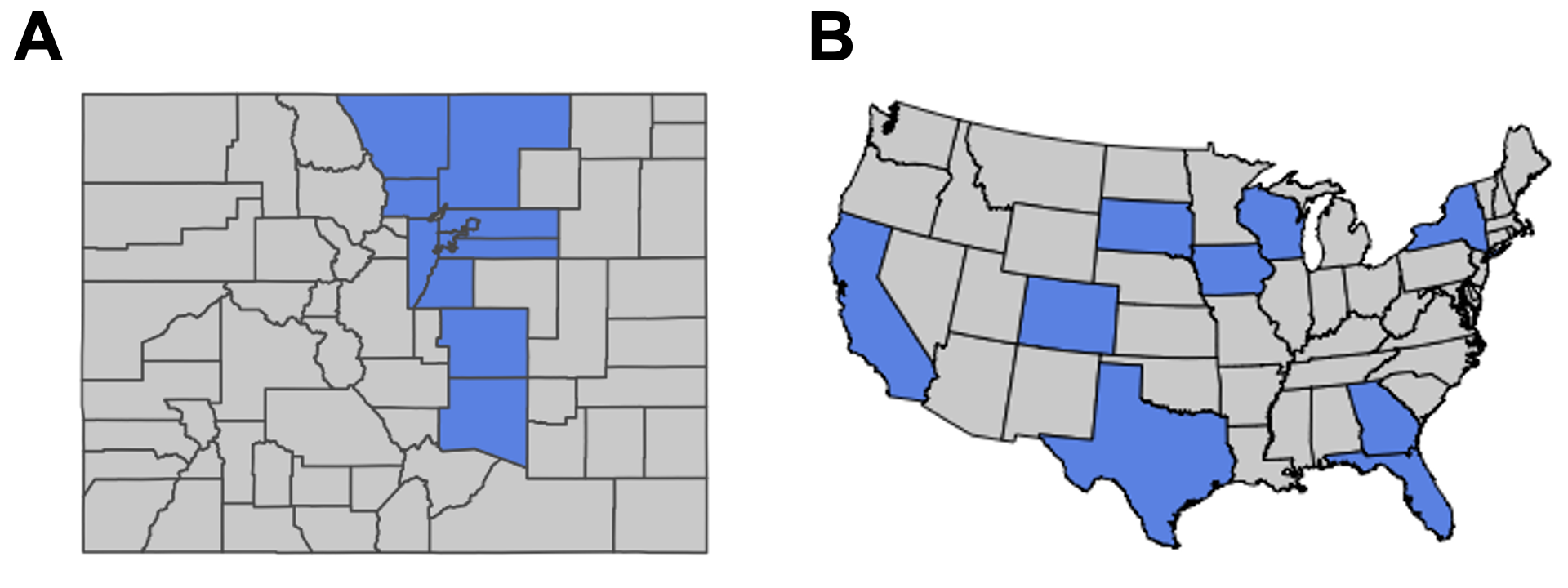}
\end{center}
	\caption{Study locations (in blue) for analysis of county-level data in Colorado (A) and state-level analysis in the United States (B).}
\label{fig:study_locations}
\end{figure}

We apply this model to county-level data in the U.S. state of Colorado (Section~\ref{sec:colorado_models}) and state-level data in the U.S. (Section~\ref{sec:states}) in the regions highlighted in Figure~\ref{fig:study_locations}. Daily counts of  cases and deaths were obtained from The New York Times\cite{TheNewYorkTimes2020}. 

\begin{figure}
\begin{center}
\includegraphics[width =0.9\textwidth]{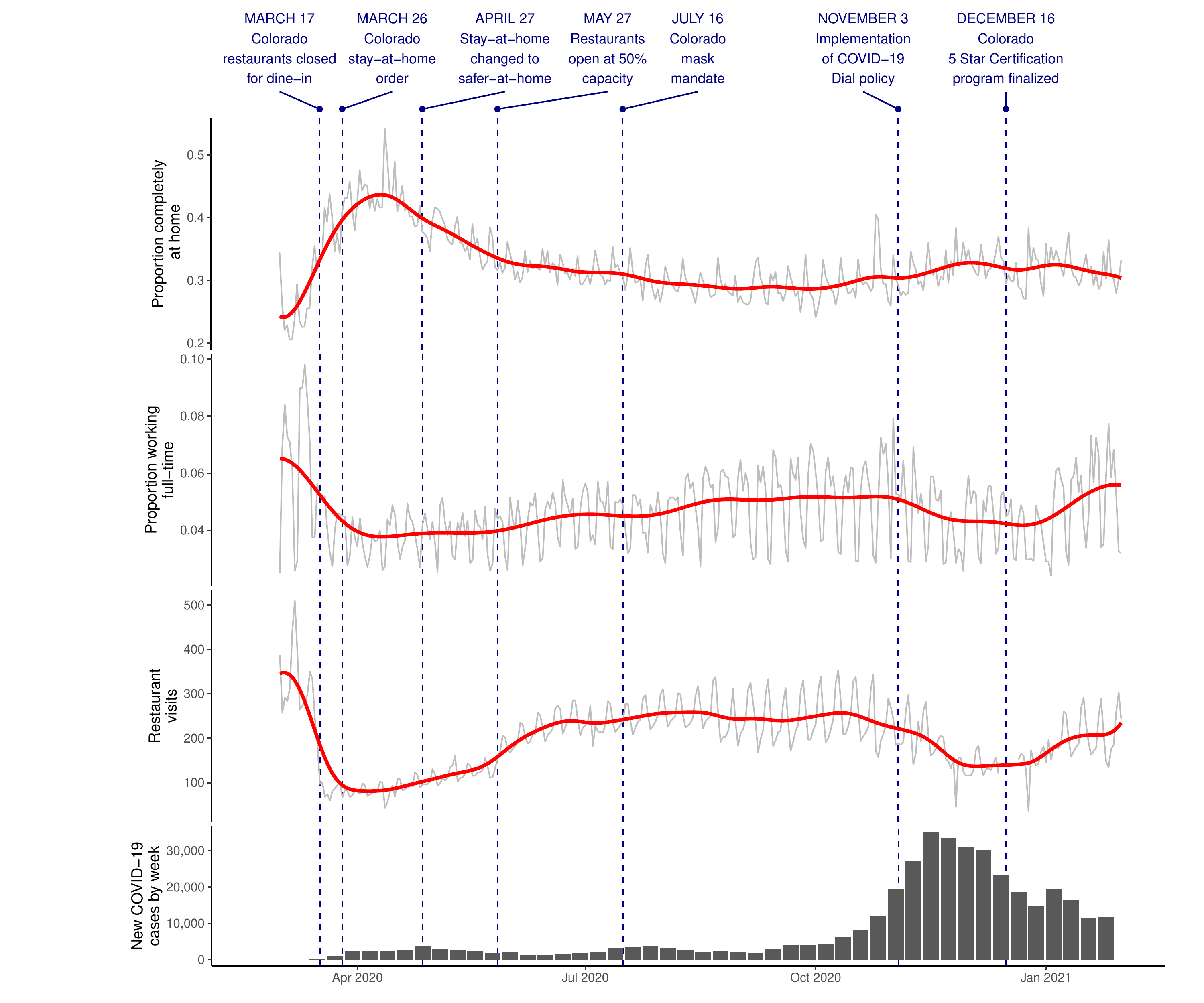}
\end{center}
\caption{Example of mobility data used in analysis. Measurements from SafeGraph were used to characterize mobility patterns, including the fraction of mobile devices that were completely at home each day, the fraction that spent 6 hours or more away from home (``full-time work''), and the number of daily visits to restaurants. This figure shows these data for the state of Colorado. The gray curves represent the 
daily values and the red curve represents the smoothed values used for modeling. Dates of several key statewide policies, as well as weekly new cases of COVID-19 in the state, are also shown.}
\label{fig:policy_timepoints}
\end{figure}

We collected mobility data from SafeGraph \cite{safegraph} via the \texttt{covidcast} R package \cite{covidcast}, which uses anonymized location data from mobile phones to generate different views of mobility over time. 
We use three mobility metrics from this data source\,---\,the fraction of mobile devices that did not leave the immediate area of their home in each day (``completely home''), the fraction of mobile devices that spent more than 6 hours at a location other than their home during the daytime (``full-time work''),  and the number of daily visits made by those with SafeGraph's apps to restaurants in a county.
 Each of these metrics are available daily, and we applied a kernel smoother to obtained smoothed values of each metric. An example of these mobility data for Colorado is shown in Figure \ref{fig:policy_timepoints}, along with the dates of some key statewide policies related to COVID-19 control. 
The mobility data for all regions is shown in Figures~S1-S3 and S15-S17 in the supplementary files.

\subsection{Models for Counties in Colorado}
\label{sec:colorado_models}
\subsubsection{Model Setup}
\label{sec:colorado_model_setup}

We first apply this model to data from counties in Colorado. The first case of COVID-19 in Colorado was reported on March 5, 2020. On March 10, 2020, the Colorado Governor issued a Declaration of Disaster Emergency and on March 27, 2020, a statewide ``Stay at Home'' order was given (Figure~\ref{fig:policy_timepoints}). Although some of the early outbreaks in Colorado were located in resort areas that have a small resident population, COVID-19 quickly spread throughout the state. On April 27, 2020, the state transitioned to a ``Safer at Home'' order that was later amended to county-specific regulations related to measures of local pandemic risk.

For modeling, we selected the ten largest counties (by population) in Colorado (Figure~\ref{fig:study_locations} and 
Table~S1 in the supplementary files). 
This excludes some of the counties with ski resort communities where the earliest outbreaks occurred, but represents the vast majority of the state's population and communities varying from urban to rural.  
For each county, we select $t=1$ as the first day with  6 or more detected cases in that county and use each day as the model time-step.
The number of initial detected cases ($I^d(0)$) was taken as the total number of reported cases in the five days before the modeling start date. We model the dynamics through January 31, 2021, at which point the widespread introduction of vaccines fundamentally changed the underlying dynamics of disease spread.

All models are fit with the priors specified in Tables~\ref{tbl:prior}, and we set $v^C = v^D = 0.1$. 
In practice, we found that the prior distributions in Table~\ref{tbl:prior_clinical} were not always informative enough for regions with a large number of cases and may lead to clinically implausible parameter estimates. Therefore, we scaled the priors in Table~\ref{tbl:prior_clinical} based on the cumulative number of detected cases.
We scale the prior standard deviations of $\alpha$, $\eta_0$, $\nu$, $\gamma_0$, $\delta_0$ and $a_{\psi}$ based on the ratio of the total number of cases of the county and that of Pueblo county. If the ratio is $r$, the prior standard deviation is scaled to $1/\sqrt{r}$ of the default prior standard deviation.

For $\beta_u(t)$, we used linear B-splines with 10 degrees of freedom (df) as the basis functions $\lambda(t)$. In addition, we included the three smoothed mobility measures in the regression component as $\theta_{\ell}(t)$ in \eqref{betau_s}. For $\omega(t)$, we used linear B-splines with 5 degrees of freedom.
In model fitting for all counties, we used 4 chains each with 1,000 warmup iterations and 1,500 post-warmup iterations. We set the maximum tree depth to 14 and \texttt{adapt\_delta} to 0.95.

\subsubsection{County Model Results}
\begin{figure}
\begin{center}
\includegraphics{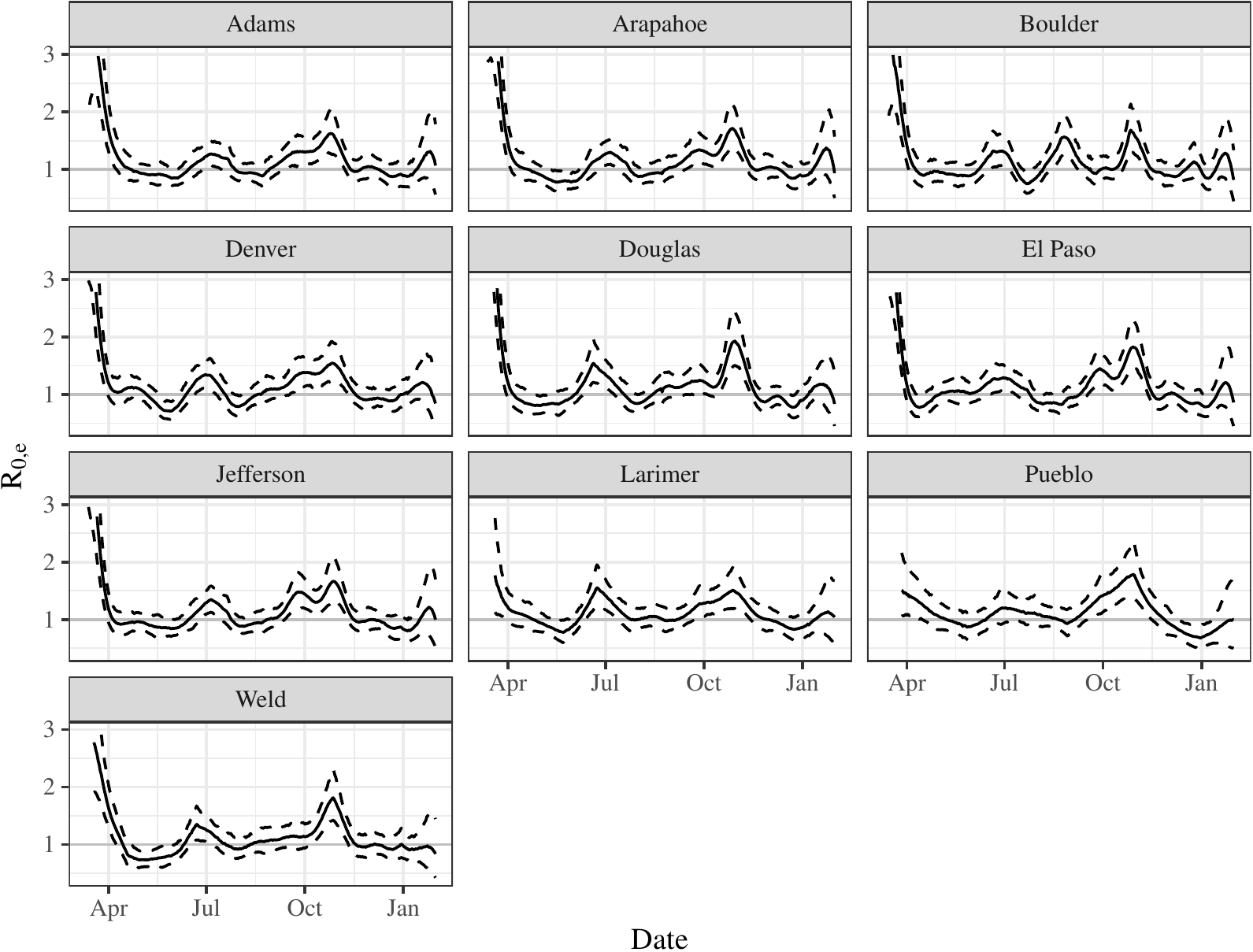}
\end{center}
\caption{Posterior mean (solid line) and 95\% credible interval (dashed line) of $\mathcal{R}_{0,e}$ for Colorado county models. Horizontal  line represents  $\mathcal{R}_{0,e}=1$.}
\label{fig:R0_CO}
\end{figure}

\begin{figure}[t]
\begin{center}
\includegraphics[scale=0.9]{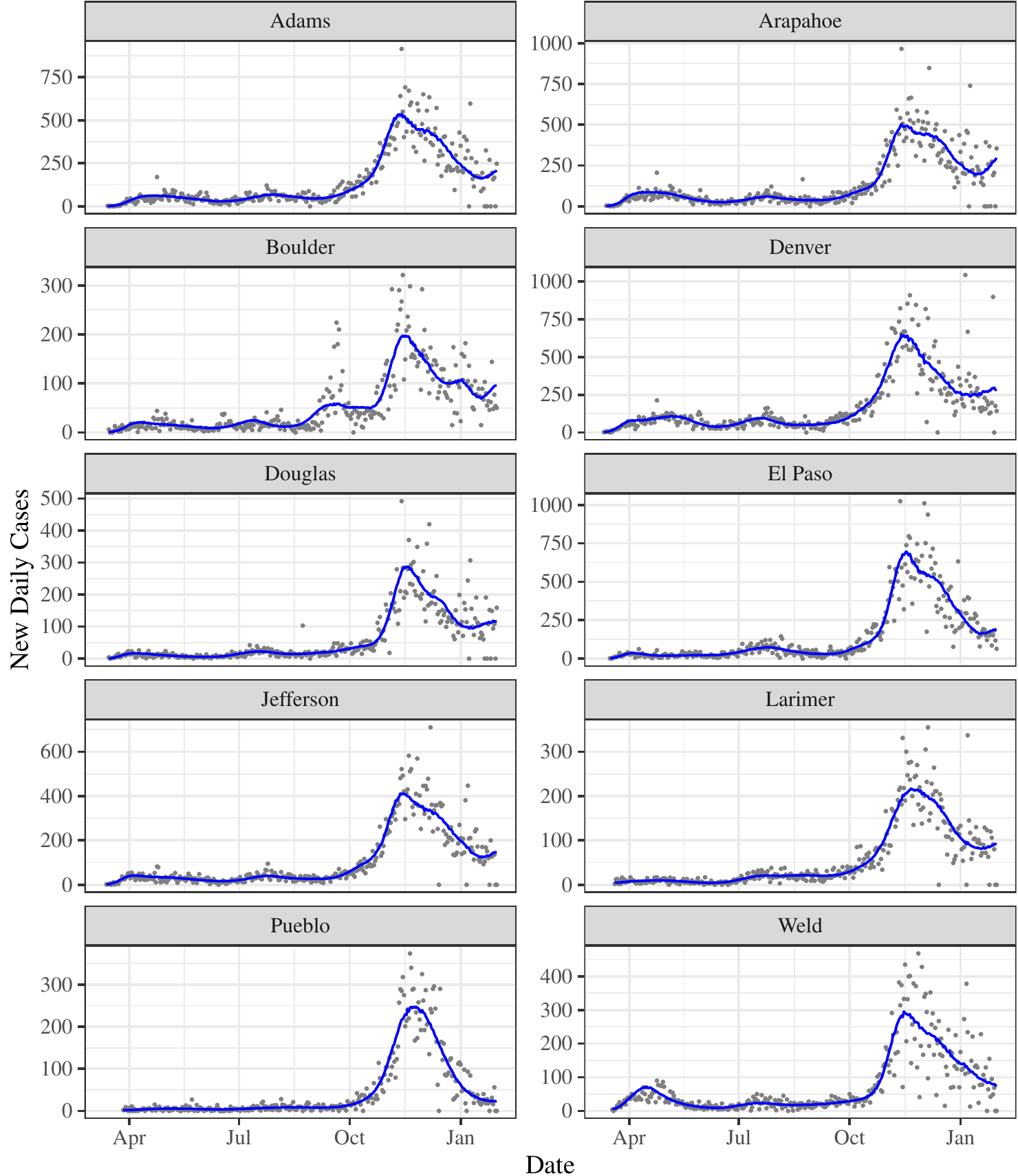}
\end{center}
\caption{New cases by date in Colorado counties. Points show observed counts, line shows average model predictions (i.e. mean of the posterior predictive distribution).}
\label{fig:newCases_CO}
\end{figure}

The time-varying $\mathcal{R}_{0,e}$ for each Colorado county is displayed in Figure~\ref{fig:R0_CO}. There is a similar tri-modal trend across all counties: peaks in $\mathcal{R}_{0,e}$ in late March (at the start of modeling), July, and November. These trends match the rise and fall of new cases, shown in Figure~\ref{fig:newCases_CO}, with the largest peak in cases happening in November and December in all counties. 
The dynamics of some counties are slightly different, such as additional peaks in $\mathcal{R}_{0,e}$ in early Fall 2020.

The impact of the stay-at-home orders is clearly evident in the decline in $\mathcal{R}_{0,e}$ throughout April and the resulting downturn of new cases in late April and early May. The equilibrium reproductive number remained near 1 until a small rise in early July. In early September, the $\mathcal{R}_{0,e}$ began to increase in the state overall, with a particularly large rise in Boulder County, where there were several high-profile outbreaks among college-aged individuals (see corresponding panel in Figure~\ref{fig:newCases_CO}). As the number of new daily cases decreased in Boulder County, the estimated $\mathcal{R}_{0,e}$ declined again.

The predicted and observed number of daily deaths is provided in Figure~S4 in the supplementary files. On most days, no deaths were reported, but the predicted number of new deaths clearly shows the  April and November peaks in mortality.

Posterior means for other parameters are shown in Figure~S5 to S12 in the supplementary files. The proportion of detected infected individuals for each day is shown in Figure~S13. 
In all counties, this proportion starts small but increases through April and May and remains elevated throughout the rest of the pandemic, reflecting increased testing capacity.

\subsubsection{Impact of Mobility on Within-County Transmission}
\begin{figure}
\begin{center}
\includegraphics[scale=0.9]{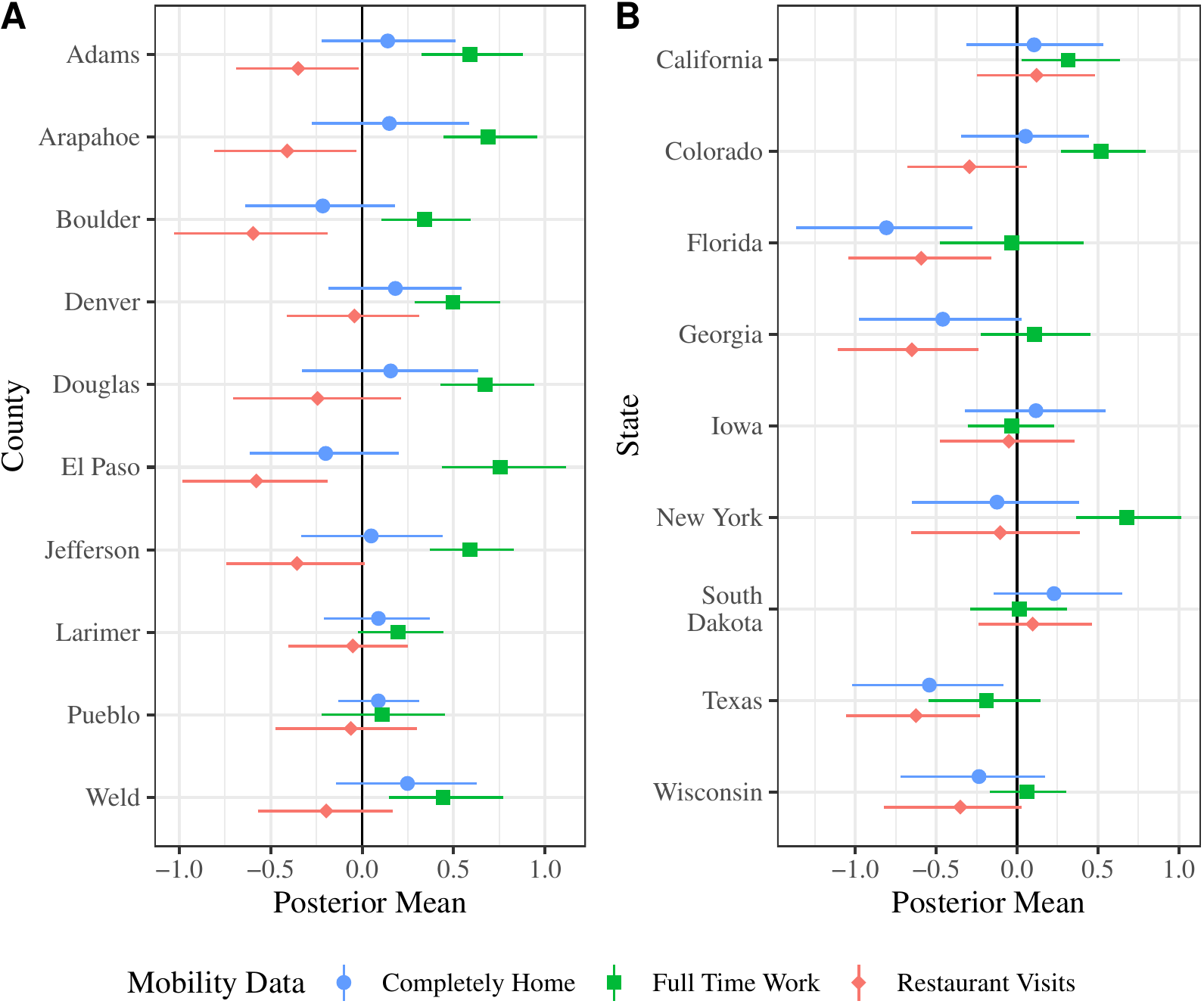}
\end{center}
\caption{Impact of mobility measures in transmission parameter  $\beta_u(t)$.  Points and horizontal error bars represent posterior means and 95\% credible intervals of the mobility coefficients,  respectively.}
\label{fig:mobilitycoef_both}
\end{figure}

\begin{figure}
\begin{center}
\includegraphics[scale=0.9]{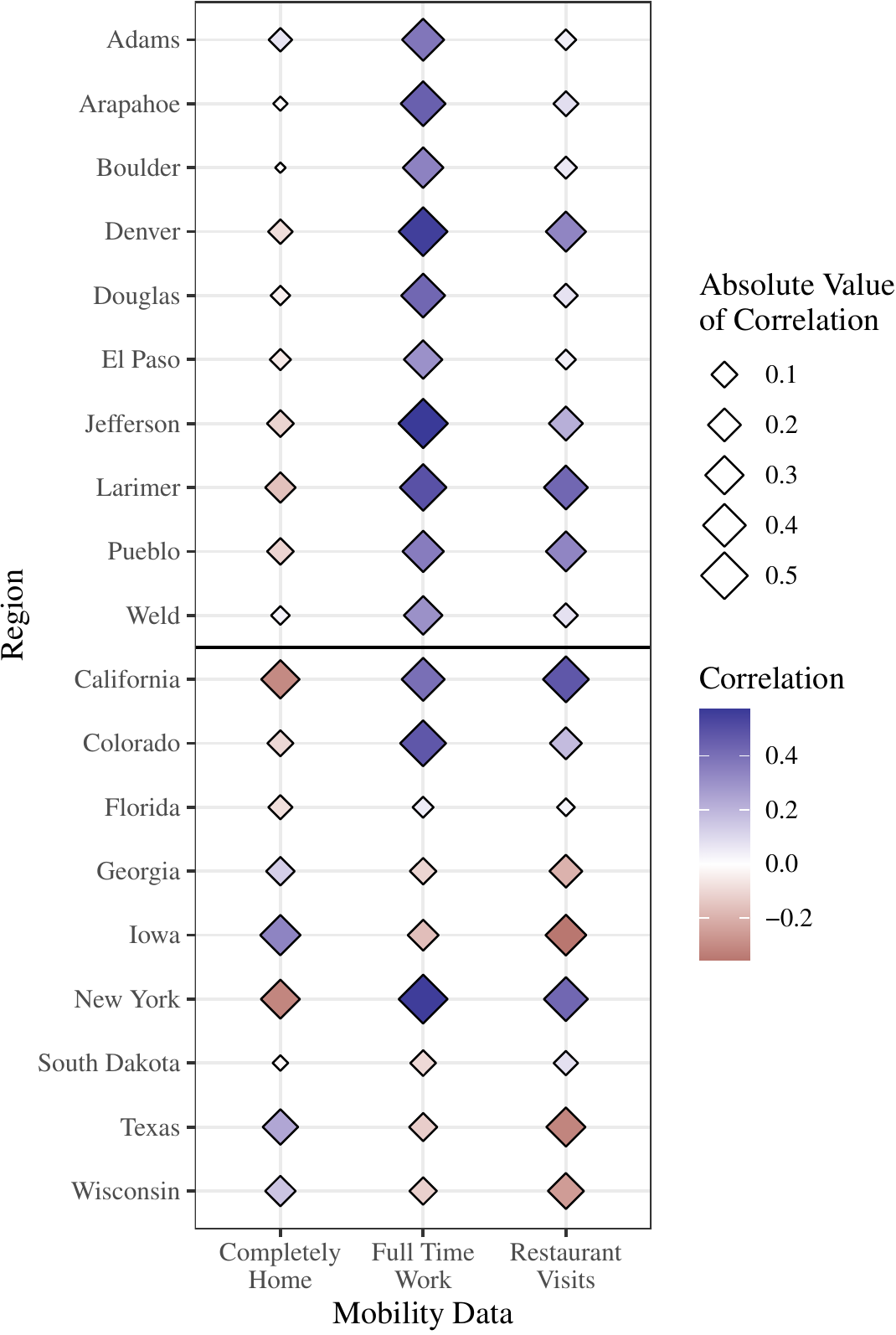}
\end{center}
\caption{Correlation of mobility data time series  with the posterior mean of transmission parameter  $\beta_u(t)$.}
\label{fig:mobilitycorr}
\end{figure}

\begin{figure}
\begin{center}
\includegraphics[scale=0.9]{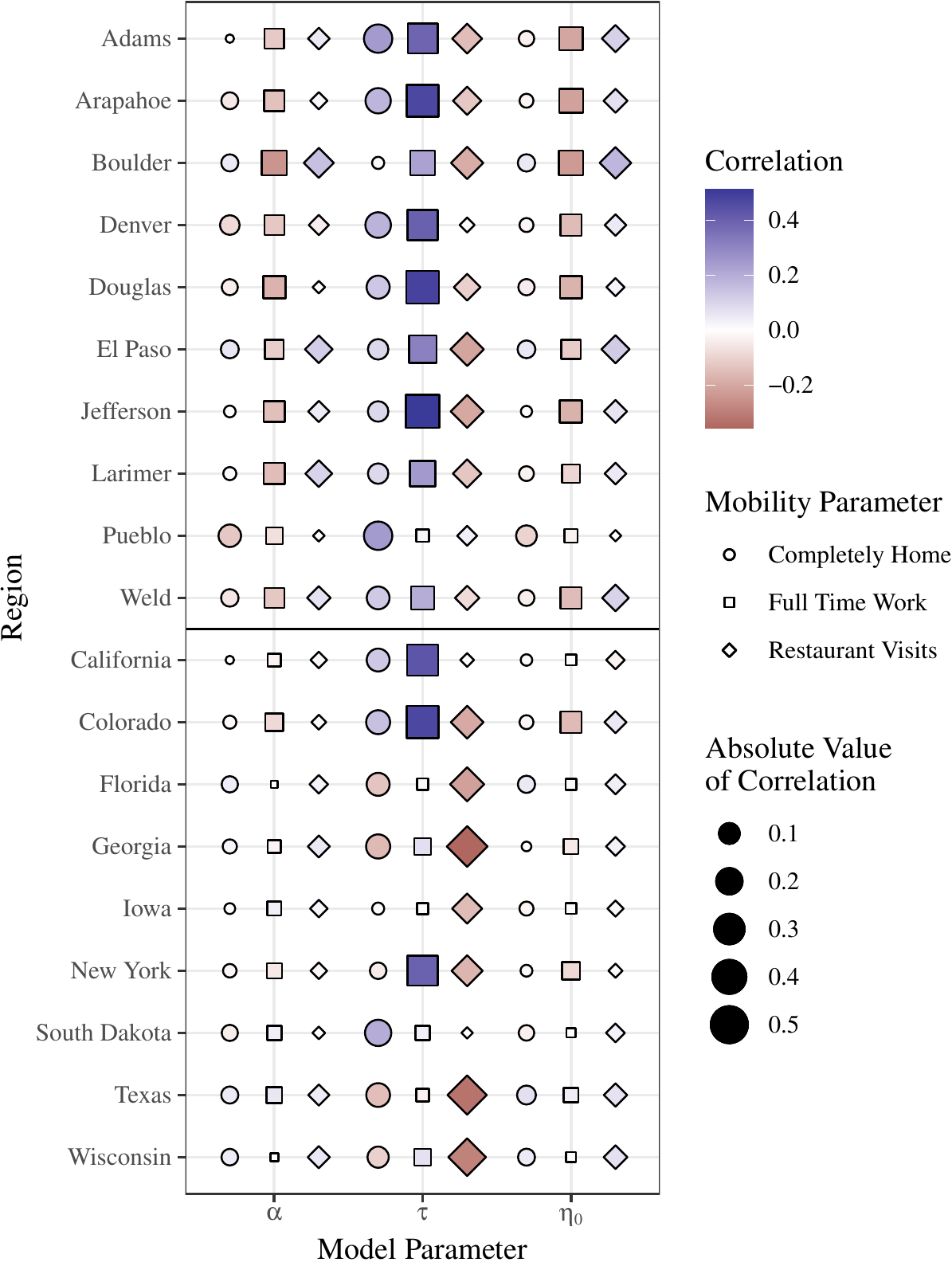}
\end{center}
\caption{Correlation of mobility data coefficients and three epidemiological parameters:  $\alpha$ (inverse of latent period), $\tau$ (reduction in transmission rate among detected infectious individuals), and $\eta_0$ (inverse of infectious period before detection).}
\label{fig:mobilitycorr_epi}
\end{figure}

County-wide mobility played an important role in modeling the transmission rate ($\beta_u(t)$) in most counties. The effects of mobility on transmission are summarized in four ways. Figure~\ref{fig:mobilitycoef_both} shows the posterior means (and 95\% credible intervals) for the coefficients  $\xi_{\beta \theta, \ell}$ for each mobility term. Figure~\ref{fig:mobilitycorr} shows the correlations between the mobility data time series and the time series of the posterior mean of $\beta_u(t)$, i.e. $\mathrm{Corr}(\theta_\ell(t), \text{E}[\beta_u(t) | \{Y^C(t), Y^D(t)\}])$.  Figure~\ref{fig:mobilitycorr_epi} shows the posterior correlation between the mobility coefficients ($\xi_{\beta \theta, \ell}$) and three of the time-constant epidemiological parameters, i.e. $\mathrm{Corr}(\xi_{\beta \theta, \ell}, \alpha | \{Y^C(t), Y^D(t)\})$ and similarly for $\tau$ and $\eta_0$. 
Figure~S14 in the supplementary files 
shows the combined time-varying effect of mobility on infectiousness (i.e., $\sum_{\ell=1}^L \theta_{\ell}(t)  \xi_{\beta \theta, \ell}$). 

In all counties, the coefficient for the proportion of people working full time is positive, indicating a positive association between this measure of mobility and COVID spread, and in all except Larimer and Pueblo counties the credible interval excludes zero. This clearly demonstrates how changing patterns in work trends (data 
displayed in Figure~S2) 
are associated with transmission of COVID-19. 
This reflects not only the sharp reduction in the proportion of people working full time after the stay-at-home orders at the start of the pandemic in March 2020,
 but also the gradual increase in mobility due to more people working away from home after the April 27 ``Safer-at-Home''  order that relaxed some restrictions through the summer and fall (Figures \ref{fig:policy_timepoints} 
 and S2). 
  The counties with the largest coefficients are all in the metropolitan areas surrounding the economic centers of Denver, Boulder, and Colorado Springs.
 Unsurprisingly, the large positive values of the mobility coefficient lead to strong positive correlations between the proportion of people working full-time and the posterior mean transmission rate (Figure~\ref{fig:mobilitycorr}). In addition, for most counties, we observe strong positive posterior correlations between the coefficient for full time work and $\tau$ while the posterior correlations between the coefficient for full time work and $\alpha$ and $\eta_0$ tend to be negative (Figure~\ref{fig:mobilitycorr_epi}).


The direct effect of visiting restaurants on the transmission rate was weaker in most counties, as evidence by the smaller (in magnitude) posterior means compared to the full-time work coefficient (Figure~\ref{fig:mobilitycoef_both}). In all counties the posterior mean of the coefficient for restaurant mobility was negative. This is also reflected in the correlation of the restaurant coefficient with the epidemiological parameters (Figure~\ref{fig:mobilitycorr_epi}), which show the opposite pattern of the full time work coefficient.
However, this must be viewed in light of concurrent adjustment for the proportion of individuals working from home (which is positively correlated with restaurant visits). When we examine the marginal correlation of restaurant mobility with transmission (Figure~\ref{fig:mobilitycorr}), we see that the relationships range from weakly positive to strongly positive.  This indicates that restaurant mobility and full-time work together are related to increases in transmission, although each mobility measure alone cannot fully explain the temporal trend in transmission. 

When concurrently adjusting for full-time work and restaurant visits, we did not see a meaningful direct impact of the proportion of individuals staying at home (credible intervals in Figure~\ref{fig:mobilitycoef_both} all include zero). However, it is important to note that this does not mean that stay-at-home orders were ineffective, but rather the mobility measures of people outside the home (at work and restaurants) were more predictive of transmission than the numbers of people staying home. The negative marginal correlation between the proportion of people at home and the posterior mean transmission rate (Figure~\ref{fig:mobilitycorr}) reflects this, indicating that more people staying home was correlated with lower disease transmission.

\subsection{Models for U.S. States}
\label{sec:states}
\subsubsection{Data and Model Setup}

We also fit the model to nine U.S. states that are representative for their different trends in the number of cases throughout the pandemic.  The selected states were (Figure~\ref{fig:study_locations}): Colorado, South Dakota, and Wisconsin, which had small early waves and a large late wave; California, Florida, and Texas, which had large numbers of cases mid-summer and a late wave; Iowa, which had moderate numbers followed by a late wave; Georgia, which had relatively small summer and fall peaks; and New York, which had a very large early peak in the spring and a late wave in November.

For these models, we chose the initial time point to be the first day after the state-level case count exceeded 100 cases (Table~S2 in the supplementary files) and modeled dynamics through January 31, 2021.  
Prior specification and model structure was the same as in the county-level models described in Section~\ref{sec:colorado_model_setup}. 
For the state-level models, we included the same three measures of mobility, but calculated at the intra-state level (Figures~S15 to S17), in the $\beta_u(t)$ term. For the clinical parameters, the prior standard deviations were scaled against the number of cases in South Dakota.

\subsubsection{State Model Results}
\begin{figure}
\begin{center}
\includegraphics{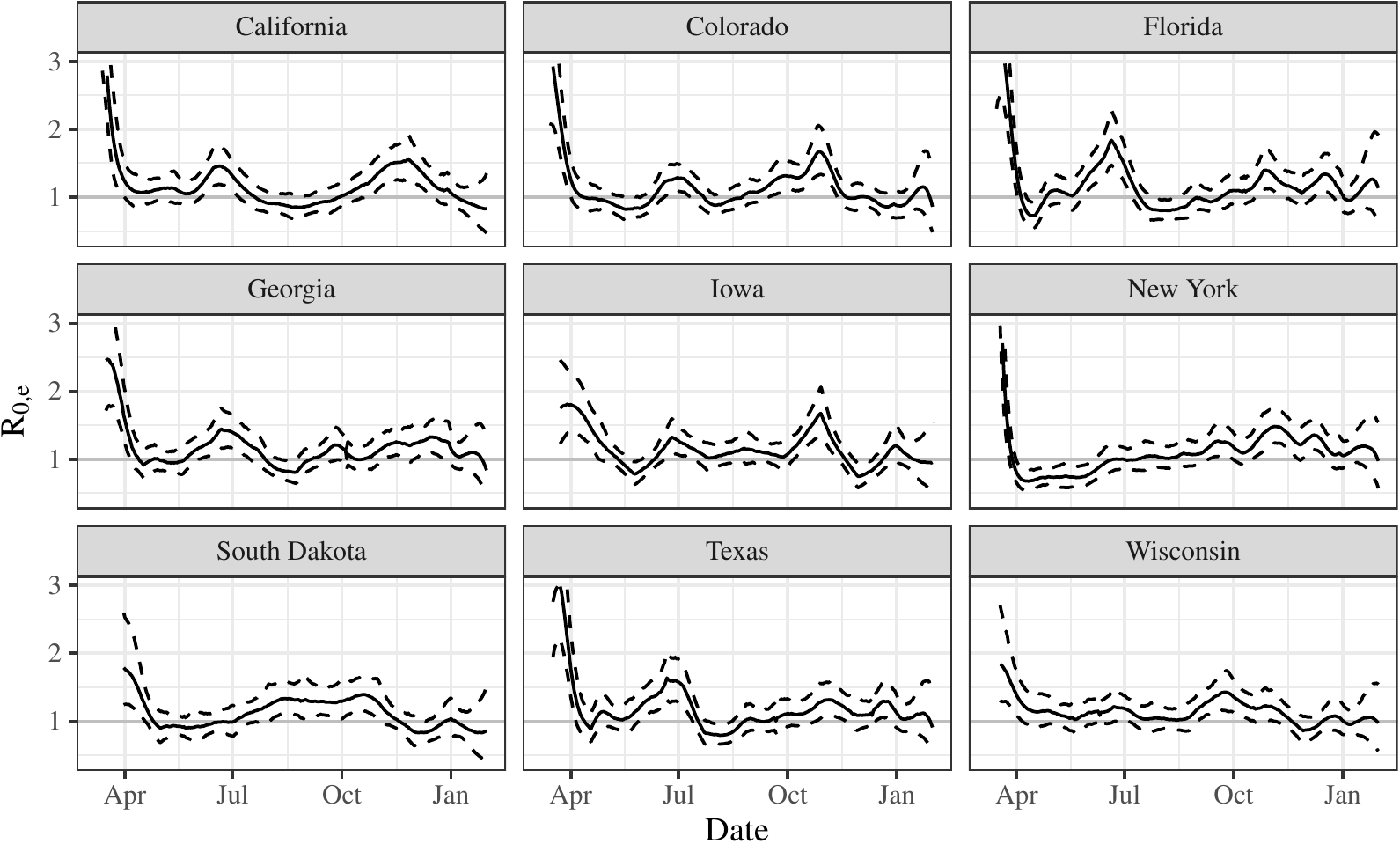}
\end{center}
\caption{Posterior mean (solid line) and 95\% credible interval (dashed line) of $\mathcal{R}_{0,e}$ in state models. Horizontal line represents  $\mathcal{R}_{0,e}=1$.}
\label{fig:R0_States}
\end{figure}

The value of $\mathcal{R}_{0,e}$ for each state is displayed in Figure~\ref{fig:R0_States} and the observed and predicted numbers of new cases for each state are plotted in Figure~S18 in the supplementary files. 
 There is a similar tri-modal trend across all states: peaks in $\mathcal{R}_{0,e}$ in April, July, and November--December.  New York began with a very large $\mathcal{R}_{0,e}$, that precipitously declined as the epidemic in New York City was brought under control in April 2020. After the initial spread of COVID-19 spurred a round of lockdown measures, most states had $\mathcal{R}_{0,e}$ values around 1 before later local peaks in $\mathcal{R}_{0,e}$ lead to additional waves of COVID-19. California, Florida, and Texas all had increases in $\mathcal{R}_{0,e}$ in late June that preceded large numbers of cases in July 
(Figure~S18 in the supplementary files). 
 Meanwhile, South Dakota had an $\mathcal{R}_{0,e}$ above 1 from July through October, which led to continuous rate of growth in new cases during that time 
 (Figure~S18).
   Consistent with the county-level models, the full-state Colorado model had its largest peak  $\mathcal{R}_{0,e}$ in November.
 
\subsubsection{Impact of Within-State Mobility}
For some states, including the overall Colorado model, there is a strong relationship between one or more of the mobility factors and the transmission parameter $\beta_u(t)$ (Figure~\ref{fig:mobilitycoef_both}). The models for California, Colorado, and New York all have coefficients for the full-time work mobility that are positive, indicating that the transmission rates are positively associated with the proportion of individuals being away from home more than six hours per day. The large coefficient for New York reflects the link between the drop in mobility in March and April 2020
 (see Figure~S28 in the supplementary files) 
 and the drop in cases after an initial large peak during that time. This impact is further reflected in the correlation between the proportion of people working from home and the transmission term (Figure~\ref{fig:mobilitycorr}).
Similar to the Colorado county models, the coefficient for full-time mobility in these three states was highly positively correlated with $\tau$ (Figure~\ref{fig:mobilitycorr_epi}). 
The impact of people working away from home full time was limited in the other state models, as evidenced by the credible intervals including zero (Figure~\ref{fig:mobilitycoef_both}) and the weak correlations with transmission (Figure~\ref{fig:mobilitycorr}). It is possible that the mobility data provide a better reflection of potential for contacts in some states compared to others. In a state with heavy tourist traffic, for example, the mobility data for working full-time or at home would not capture tourist patterns, although the restaurant visits would to some degree. This may partially explain why a weaker association was observed in states like Florida and South Dakota. 

The impact of restaurant mobility was heterogeneous across states. In California, Colorado, and New York, the coefficient was not meaningfully different from zero (Figure~\ref{fig:mobilitycoef_both}) but restaurant mobility was positively correlated with transmission (Figure~\ref{fig:mobilitycorr}). The models for Florida, Georgia, and Texas yielded negative posterior means for the coefficients of restaurant mobility. However, this must be viewed in conjunction with the negative posterior means for the coefficient of the proportion of individuals completely home (for which a negative value means less disease spread when more people are staying home). 
The net result was a minimal effect of mobility on disease spread in those particular states 
(Figure~S28).

Overall, the correlations between the time-varying transmission rate and the proportion of people completely home were stronger among states than in the Colorado county models. This was particularly true in California and New York, where the increase in the proportion of individuals working from home was particularly large 
(Figure~S15). 
 However, this correlation was largely captured by the spline terms in $\beta_u(t)$, and so the coefficients for the proportion of people completely at home were not meaningfully different from zero (Figure~\ref{fig:mobilitycoef_both}) despite a negative correlation between this measure of mobility and transmission (Figure~\ref{fig:mobilitycorr}).
This reflects a challenge of the flexibility of our modeling approach: having both temporal splines and time-varying predictors means that variation can be explained in multiple ways. In the case of the California and New York models, the correlation between transmission and proportion of people completely at home is likely driven by the dramatic effects of stay-at-home orders at the beginning of the pandemic, while the later surges in transmission (Figure~\ref{fig:R0_States}) do not correlate as well with the proportion of people completely at home. This leads to the temporal variability being attributed to the temporal splines, rather than the mobility time series.

In Iowa and Texas, the correlation between transmission and mobility measures was opposite from what was seen in the majority of models. In these two states, more people working from home was positively correlated with transmission and more people visiting restaurants was negatively correlated with transmission (Figure~\ref{fig:mobilitycorr}). These surprising correlations may be driven in part by the reduction in mobility around the time of the December transmission waves 
(compare Figure~\ref{fig:R0_States} and Figure~S16 in the supplementary files).
 Or it may be that differences in policies (such as mask-wearing) mean that mobility alone is not representative of transmission risk in these states.

The correlation between mobility data coefficients and $\alpha$ and $\eta_0$ were weaker in the state models that in the Colorado county models (Figure~\ref{fig:mobilitycorr_epi}). 
At the state level, the larger amount of case data means that $\alpha$ and $\eta_0$ can be estimated with more precision
 (compare Figures~S7 and S21 in the supplementary files), 
 reducing the connection between mobility and these clinical parameters.  However, the stronger correlation between mobility coefficients and $\tau$, which directly affects transmission, remains.

\section{Conclusions and Discussions}
\label{sec5}
We have presented a flexible model for a time-varying infectious disease that includes: temporal variation in key model parameters through the use of splines and a regression term, a multiplicative state-space process model for allowing for day-to-day heterogeneity in disease spread, and overdispersion in the observed number of cases. Together, these elements allow our model to capture important features of data observed in a pandemic. 

The inclusion of the mobility term in the infectiousness parameter $\beta_u(t)$ was of particular interest to us. We observed that it played a role in the transmission in most counties and states that we modeled (Figure~\ref{fig:mobilitycoef_both}). However, it did not play a role in the modeled transmission in all regions. The heterogeneous impact of mobility data may reflect the importance of other factors, such as compliance with mask-wearing mandates. 
While derived from individual-level information on movement, the mobility data we were able to incorporate into the models was averaged at the county and state levels. This relatively coarse spatial scale means that we cannot differentially model the connection between disease transmission and mobility in at-risk or highly-active subpopulations. Individual-level information on movement could improve the strength of the evidence between mobility and COVID transmission.
Furthermore, a potential downside to using cell phone movement as a proxy for human mobility is that there are disparities in cell phone ownership and usage among groups of people. In particular, older Americans are less likely to own a cell phone, and thus may be left out of the mobility discussion. Similarly, cell phone use may also be affected by socioeconomic status and not capture institutionalized populations well.
Despite these limitations of the mobility data, we saw a clear effect of reduced mobility in the sharp reduction in cases early in the pandemic (Figure~\ref{fig:R0_CO}). Furthermore, higher rates of working outside the home were associated with increased transmission in many counties and states.

The multiplicative process model introduces extra flexibility to the compartmental model, which better reflects fluctuations in the rates of detection and transmission. In addition to non-random mixing of the population, a key violation of the traditional SIR model assumptions is that each individual is equally infectious. There is now considerable evidence \cite{althouse2020superspreading, adam2020clustering, lau2020characterizing,
lemieux2021phylogenetic} that superspreaders play an important role in the spread of COVID-19, and the flexibility afforded by the process model can incorporate this heterogeneity in spread by allowing for a multiplicative shift to the solution to differential equations of the compartmental model at each time point. 
This has the additional advantage of allowing the process to explicitly account for heterogeneity in case reporting, such as the tendency of many states to report fewer cases on weekends. 

Our approach also presents advantages over approaches that only compare time series of cases and other factors to identify marginal relationships \cite{kraemer2020effect,glaeser2020much, iacus2020human}.  By incorporating information on mobility into a model for disease spread, our model can more accurately represent a potentially causal role of factors in disease spread. This further allows for comparison of mobility information with the latent rate of disease transmission, rather than just the number of reported cases.

One general challenge in the fitting and interpretation of the compartment model is the limited data for the number of individuals at each stage of the dynamics. For our model, although the framework presented in Section~\ref{sec:math_model} contains eight compartments, we only observe data from two of those compartments ($I^d$ and $D^d$). Several parameters influence the number of new cases $\mu^C(t)$, which can lead to some identifiability issues. 
For example, a moderate number of identified cases could occur due to a low detection (small $\psi(t)$) among a large number of undetected infectious individuals (large $I^u$, arising from large $\beta_u(t)$), or it could be due to high detection (large $\psi(t)$) among a small number of infectious individuals (smaller $I^u$, arising from smaller $\beta_u(t)$).
To mitigate this, we introduced a non-decreasing structure for $\psi(t)$. Nonetheless, it remains nontrivial to accurately estimate the true number of undetected cases at any given time and the 
results of Figure~S13
should be interpreted cautiously. Inclusion of information on testing rates could be used to partially address this by providing a basis for $\psi(t)$, but quality and quantity of information on testing rates and strategies vary widely between and within jurisdictions, which deserves more modeling efforts in the future.

Several extensions are possible to the models we presented here. The additive model framework for the time-varying parameters could be expanded to include environmental factors such as temperature, humidity, and air pollution. 
There has been suggestive evidence \cite{bashir2020correlation, frontera2020severe, fronza2020spatial,sajadi2020temperature, shi2020impact, travaglio2021links, wu2020effects,  zhu2020association} that these factors may influence spread and severity of symptoms. As noted above, more detailed mobility information could be incorporated to represent small-scale movement patterns. 
Although we used the same prior distributions and model settings across all counties and states, a natural extension of this approach would be to fit a joint model that includes all regions (counties or states) together. 
This would allow for information on the non-spatial parameters (e.g. $\alpha$ and $\nu$) to be estimated using shared information. Direct movement of susceptible and infectious cases between regions could then be included as well.
The primary drawback to such an approach is the computational complexity that arises from having separate spline parameters in each region.

Overall, the proposed model provides a rigorous, process-driven framework for modeling the impact of time-varying factors on  infectious disease spread. Our analyses showed an important role of mobility in the spread of COVID-19 in several Colorado counties and U.S. states.

\section{Data Accessibility}
The data and R code used to fit these models is available at \url{https://github.com/covid19-csu/covid-model-data}.



\bibliographystyle{apacite}

\end{document}